\documentclass{mem}
\usepackage{graphicx}
\idline{75}{282}
\begin{document}
\def\teff{$T\rm_{eff }$}
\def\kms{$\mathrm {km s}^{-1}$}

\title{Classical Cepheids as Age Indicators
}

   \subtitle{}

\author{
M. \,Marconi\inst{1} 
\and G. \, Bono\inst{2}
\and F. \, Caputo\inst{2}
\and S. \, Cassisi\inst{3}
\and P. \, Pietrukowicz\inst{4}
\and G. \, Pietrzynski\inst{5}
          }

  \offprints{M. Marconi}

\institute{
INAF -- Osservatorio Astronomico di Capodimonte, Via Moiariello 16,
80131 Napoli, Italy, \email{marcella@na.astro.it}
\and
INAF -- 
Osservatorio Astronomico di Teramo, Via Maggini 47,
64100 Teramo, Italy
\and
INAF -- 
Osservatorio Astronomico di Roma, Via Frascatii 33,
00040 Monte Porzio Catone, Italy
\and
Nicolaus Copernicus Astronomical Centre, ul. Bartycka 18, 
00716 Warsaw, Poland
\and
Universidad de Conception -- Dept. de Fisica, Casilla 160-C, Concepción, C
Chile
}

\authorrunning{Marconi}

\titlerunning{Classical Cepheids as Age Indicators}

\abstract{
Theoretical Period-Age and Period-Age-Color relations for different  
chemical compositions have been recently derived using an updated
homogeneous set of evolutionary and pulsational models. We apply  
these relations to Cepheids in the Magellanic Clouds to constrain 
the recent star formation history of these dwarf galaxies. 
Finally, we also compare the radial distribution of classical 
Cepheids with young clusters and star forming regions.  
}
\maketitle{}

\section{Introduction}

Classical Cepheids are traditionally used as primary distance
indicators, since their pulsation periods, colors, and intrinsic 
luminosity are tightly connected. Moreover, they are good tracers of 
intermediate-mass stars in the Galactic disk (Kraft \& Schmidt 1963), 
and star-forming regions in extragalactic systems (Elmegreen \& Efremov
1996). The use of Cepheids as tracers of young 
stellar population was soundly supplemented by the evidence that if 
these objects obey to a PL relation, and to a Mass-Luminosity (ML) 
relation, they also obey to a Period-Age (PA) relation. 
In particular, an increase in period implies an increase in 
luminosity, i.e. an increase in the  stellar mass, and in turn a
decrease in the Cepheid age. On the basis of these plane physical
arguments  several empirical and semiempirical  PA
relations for Galactic, LMC, and M31 Cepheids have been derived
to constrain the recent 
star formation history of the explored regions (Efremov 1978;  
Magnier et al. 1997;   Efremov \& Elmegreen 1998; Grebel \& Brandner 1998; 
Efremov 2003).  The PA relation presents the following advantages: 
i) age estimates rely on the pulsation period, an observable marginally 
affected by systematic errors; 
ii) it can be applied to individual objects; 
iii) the application to cluster Cepheids provides a unique opportunity 
to evaluate the age of the parent cluster,  even if the photometry of 
Turn-Off stars is lacking or poor. However, current PA relations are
calibrated using cluster Cepheids. This means that the slope and the
zero-point of the PA relation might be affected by distance and reddening
uncertainties, as well as by the period range covered by cluster Cepheids. 
Finally, age estimates based on the PA relations rely on the assumption 
that the Cepheid instability strip has a negligible width in temperature. 
This working hypothesis introduces systematic errors for periods longer 
than $\sim$ 8 days. In order to overcome this problem {\it pulsational} ages 
should account for the color dependence by using a Period-Age-Color (PAC) 
relation. To provide a new theoretical framework we constructed detailed 
sets of evolutionary tracks and nonlinear convective pulsation models 
covering a broad range of stellar masses and chemical compositions 
(Bono et al. 2005). On the basis of these calculations we derived new 
PA and PAC relations both for fundamental and first overtone pulsators.
We performed a detailed comparison between evolutionary and pulsation 
ages for a sizable sample of LMC (15) and SMC (12) clusters and for 
two Galactic clusters (Bono et al. 2005). As a result, we found that 
the different age estimates agree at the  level of 20\% for LMC and 
Galactic clusters and of 10\% for SMC clusters.
These findings support the use of PA and PAC relations to supply
accurate  estimates of individual stellar ages in the Galaxy and in
external Galaxies.  In this paper we extend the analysis to field 
Cepheids in the Magellanic Clouds.

\section{PA and PAC relations}

The sample of Magellanic Cepheids is based on the huge database collected by 
the OGLE project (Udalski et al. 1999a,b) together with the old Payne-Gaposchkin 
catalogue (see Figs. 1 and 2) for long-period objects. In order to evaluate the 
individual ages of selected Cepheids we applied predicted PA and V-I color 
PAC relations for Z=0.008, Y=0.25 (LMC) and Z=0.004, Y=-0.25 (SMC) to 
fundamental and first overtone pulsators. For the OGLE samples we derived 
pulsation age estimates by averaging the results obtained from the PA and 
the V-I  PAC relations, while for the Payne-Gaposchin sample both the 
colors and the pulsation mode are not available and we used the fundamental 
PA relation.

\begin{figure}[]
\resizebox{\hsize}{!}{\includegraphics[clip=true]{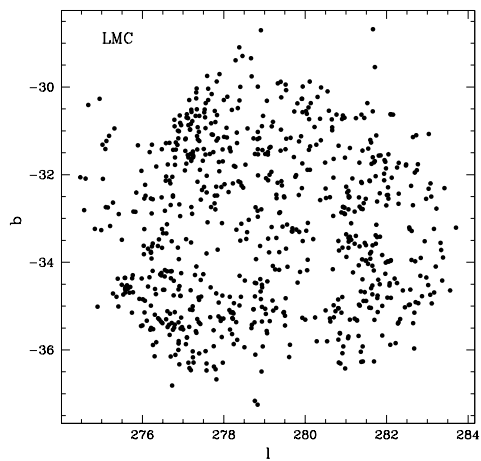}}
\caption{\footnotesize Distribution in Galactic coordinates of the 
Payne-Gaposchin LMC Cepheids.
}
\label{fig1}
\end{figure}

\begin{figure}[]
\resizebox{\hsize}{!}{\includegraphics[clip=true]{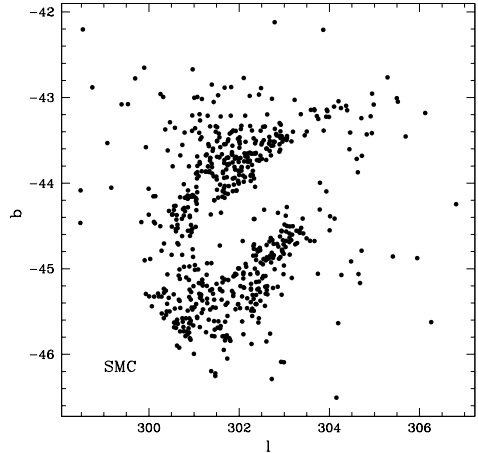}}
\caption{ \footnotesize Distribution in Galactic coordinates of the 
Payne-Gaposchin SMC Cepheids.
}
\label{fig1}
\end{figure}


In the left panels of Figs. 3, 4, and 5 we show the Cepheid distributions in 
Galactic coordinates for LMC (top) and SMC (bottom) objects and stellar ages 
ranging from less than 20 to more than 80 Myr. The right panels show the same 
distribution but for stellar clusters with age estimates based on integrated 
photometry or CMD properties  (Pietrzynski \& Udalski 1999, 2000; 
Efremov 2004; Girardi et al. 1995; Mackey \& Gilmore 2003a,b). As expected,  
the radial distribution of very young Cepheids ($t \le 20$ Myr) is peaked on 
the center of the MCs. In this age range are available a limited nuumber of 
clusters (see Fig. 3), but the two radial distributions appear quite similar. 
Young associations are also present in the MCs, but their age estimates are 
not as accurate as for star clusters. For ages ranging from 20 to 80 Myr the 
radial distribution of Cepheids present a more complex pattern when compared 
with the star clusters. The star clusters appear to be good tracers of the 
LMC bar. On the other hand, the Cepheids are good tracers not only for the 
bar but also for the star forming regions (30 Dor and LH77, see Fig. 4). They 
also display a disk like distribution in the regions located outside the bar 
that the star clusters do not show. The radial distribution of Cepheids and 
star clusters in SMC show the typical cigar-like shape. However, Cepheids 
show a spur located at $l <  300$ and $b\sim -44$ degrees that is not present 
in the star cluster distribution.   

The radial distribution of LMC Cepheids with ages older than 80 Myr presents 
a complex patter when compared with the star clusters. The latter objects trace 
very-well the LMC bar, while the Cepheids show several off-center peaks outside 
the bar. One of this peaks ($l\sim 277$ and $b\sim -31$) is associated with  
30 Dor, thus suggesting the ongoing star formation activity in this region for 
at least 60 Myr. The radial distribution of the SMC with ages older than 80 Myr 
is quite similar to the distribution of the star clusters. The larger extent 
in longitude might only be due to selection effects.

\begin{figure}[]
\resizebox{\hsize}{!}{\includegraphics[clip=true]{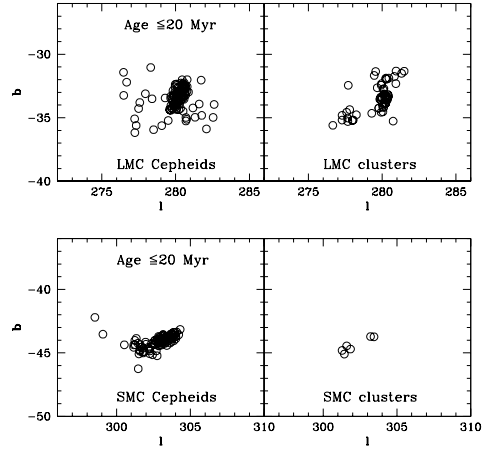}}
\caption{ \footnotesize Cepheid versus clusters distributions for ages younger 
than 20 Myr in the LMC (top) and in the SMC (bottom).
}
\label{fig3}
\end{figure}

\begin{figure}[]
\resizebox{\hsize}{!}{\includegraphics[clip=true]{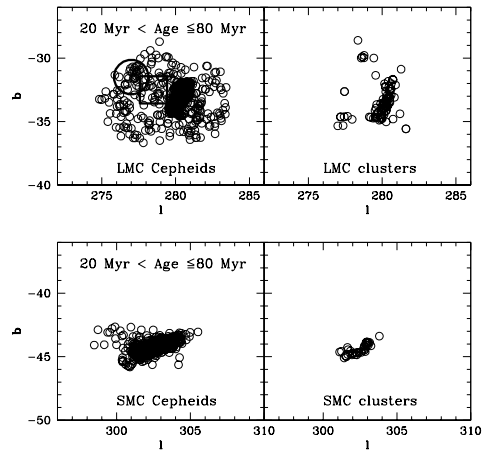}}
\caption{ \footnotesize Same as in Fig. 3, but for ages ranging from 20 to 80 Myr. 
The large symbols plotted in the top left panel mark the position of two 
well-known star forming regions, namely 30 Dor (square) and LH77 (circle). 
}
\label{fig4}
\end{figure}

\begin{figure}[]
\resizebox{\hsize}{!}{\includegraphics[clip=true]{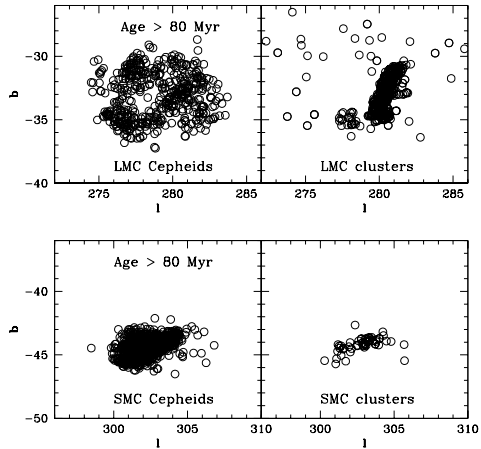}}
\caption{ \footnotesize Same as in Fig. 3, but for ages older than 80 Myr.
}
\label{fig5}
\end{figure}

\section{Conclusions}
New accurate PA and PAC relations have been recently obtained on the basis of
homogeneous and updated sets of evolutionary and pulsational computations 
(Bono et al. 2005).  These relations have been applied to extended databases 
of Magellanic field Cepheids to derive individual {\it pulsational} ages. 
The radial distribution in Galactic coordinates of the investigated pulsators 
for different age bins provides useful constraints on the star formation history 
of the Magellanic Clouds. In particular, the comparison with similar results 
based on cluster ages estimated from independent methods, seems to suggest that 
Cepheid age estimates may provide a more detailed picture (higher spatial 
resolution) of recent star formation episodes in the MCs.

\begin{acknowledgements}
Financial support for this study was provided by MIUR under the scientific projects 
"Stellar Populations in the Local Group" (PI: M. Tosi) and "Continuity and 
Discontinuity in the Milky Way Formation" (PI: R. Gratton).
\end{acknowledgements}

\bibliographystyle{aa}

\end{document}